\documentclass[aps,prl,superscriptaddress,twocolumn,10pt]{revtex4-2}

\usepackage{blindtext}
\usepackage{siunitx}
\usepackage{acronym}
\usepackage{graphicx}
\usepackage{physics}
\usepackage{hyperref}
\usepackage[nameinlink, capitalise]{cleveref} 

\newacro{afm}[AFM]{atomic force microscope}
\newacro{siv}[SiV$^-$]{negatively-charged silicon-vacancy}
\newacro{nd}[ND]{nanodiamond}
\newacro{pcc}[PCC]{photonic crystal cavity}
\newacro{SiN}[Si\textsubscript{3}N\textsubscript{4}]{silicon nitride}
\newacro{cpt}[CPT]{coherent population trapping}
\newacro{eom}[EOM]{electro optical modulator}
\newacro{cqed}[cQED]{cavity quantum electrodynamics}
\newacro{ple}[PLE]{photoluminescence excitation}
\newacro{TE}[TE]{transverse electric}

\newacroplural{pcc}[PCCs]{photonic crystal cavities}

\newcommand{\siv}{\ac{siv} center\xspace}
\newcommand{\sivs}{\ac{siv} centers\xspace}
\newcommand{\nd}{\ac{nd}\xspace}
\newcommand{\nds}{\acp{nd}\xspace}
\newcommand{\pcc}{\ac{pcc}\xspace}
\newcommand{\afm}{\ac{afm}\xspace}
\newcommand{\SiN}{\ac{SiN}\xspace}
\newcommand{\cpt}{\ac{cpt}\xspace}
\newcommand{\eom}{\ac{eom}\xspace}
\newcommand{\Tone}{$T_\mathrm{1}$\xspace}
\newcommand{\cqed}{\ac{cqed}\xspace}
\newcommand{\ple}{\ac{ple}\xspace}

\newcommand{\TE}{\ac{TE}\xspace}

\newcommand{\SuMa}{see Supplemental Material \cite{SeeSupplementalMaterial}\xspace}
\newcommand{\bSuMa}{(see Supplemental Material \cite{SeeSupplementalMaterial}\xspace)}

\acresetall

\begin{document}

\title{All-Optical Spin Initialization via a Cavity Broadened Optical Transition in On-Chip Hybrid Quantum Photonics}


\author{Lukas~Antoniuk}
\altaffiliation{These authors contributed equally.}
\affiliation{Institute for Quantum Optics, Ulm University, 89081 Ulm, Germany}
\author{Niklas~Lettner}
\altaffiliation{These authors contributed equally.}
\affiliation{Institute for Quantum Optics, Ulm University, 89081 Ulm, Germany}
\affiliation{Center for Integrated Quantum Science and Technology (IQST), Ulm University, Albert-Einstein-Allee 11, 89081 Ulm, Germany}
\author{Anna~P.~Ovvyan}
\affiliation{Institute of Physics and Center for Nanotechnology, University of M\"unster, D-48149 M\"unster, Germany}
\affiliation{CeNTech — Center for Nanotechnology, 48149, M\"unster, Germany}
\author{Simon~Haugg}
\affiliation{Institute for Quantum Optics, Ulm University, 89081 Ulm, Germany}
\author{Marco~Klotz}
\affiliation{Institute for Quantum Optics, Ulm University, 89081 Ulm, Germany}
\author{Helge~Gehring}
\affiliation{Institute of Physics and Center for Nanotechnology, University of M\"unster, D-48149 M\"unster, Germany}
\affiliation{CeNTech — Center for Nanotechnology, 48149, M\"unster, Germany}
\affiliation{SoN — Center for Soft Nanoscience, 48149, M\"unster, Germany}
\author{Daniel~Wendland}
\affiliation{Institute of Physics and Center for Nanotechnology, University of M\"unster, D-48149 M\"unster, Germany}
\affiliation{CeNTech — Center for Nanotechnology, 48149, M\"unster, Germany}
\affiliation{SoN — Center for Soft Nanoscience, 48149, M\"unster, Germany}
\author{Viatcheslav N. Agafonov}
\affiliation{GREMAN, UMR 7347 CNRS, INSA-CVL, Tours University, 37200 Tours, France}
\author{Wolfram~H.~P.~Pernice}
\affiliation{Institute of Physics and Center for Nanotechnology, University of M\"unster, D-48149 M\"unster, Germany}
\affiliation{CeNTech — Center for Nanotechnology, 48149, M\"unster, Germany}
\affiliation{SoN — Center for Soft Nanoscience, 48149, M\"unster, Germany}
\affiliation{Kirchhoff-Institute for Physics, Heidelberg University, Im Neuenheimer Feld 227, 69120 Heidelberg, Germany}
\author{Alexander~Kubanek}
\email[Corresponding author: ]{alexander.kubanek@uni-ulm.de}
\affiliation{Institute for Quantum Optics, Ulm University, 89081 Ulm, Germany}
\affiliation{Center for Integrated Quantum Science and Technology (IQST), Ulm University, Albert-Einstein-Allee 11, 89081 Ulm, Germany}
\date{\today}
\begin{abstract}
Hybrid quantum photonic systems connect classical photonics to the quantum world and promise to deliver efficient light-matter quantum interfaces while leveraging the advantages of both, the classical and the quantum, subsystems. However, combining efficient, scalable photonics and solid state quantum systems with desirable optical and spin properties remains a formidable challenge. In particular the access to individual spin states and coherent mapping to photons remains unsolved for these systems. In this letter, we demonstrate all-optical initialization and readout of the electronic spin of a negatively-charged silicon-vacancy center in a nanodiamond coupled to a silicon nitride photonic crystal cavity. We characterize relevant parameters of the coupled emitter-cavity system and determine the silicon-vacancy center's spin-relaxation and spin-decoherence rate. Our results mark an important step towards the realization of a hybrid spin-photon interface based on silicon nitride photonics and the silicon-vacancy center's electron spin in nanodiamonds with potential use for quantum networks, quantum communication and distributed quantum computation.
\end{abstract}
%
%
\maketitle
%
%
Quantum applications, such as quantum networks and quantum repeaters, rely on efficient mapping of quantum information between stationary qubits and flying qubits \cite{kimbleQuantumInternet2008, wehnerQuantumInternetVision2018, rufQuantumNetworksBased2021}. While photons are a natural choice for flying qubits, since they interact weakly with their environment \cite{northupQuantumInformationTransfer2014}, numerous candidates for stationary qubits are under investigation with individual strengths and weaknesses. Hybrid quantum photonics offers a unique route to efficiently interface stationary to flying qubits, by combining individually optimized photonic and quantum systems. The \siv in diamond is of particular interest, since its defect symmetry enables to retain excellent optical properties even in diamonds with sizes below the optical wavelengths \cite{rogersSingleSiVCenters2019}, so-called \nds, enabling two-photon interference from \sivs in remote \nds \cite{waltrichTwophotonInterferenceSiliconvacancy2023}. Hybrid quantum photonics \cite{kubanekHybridQuantumNanophotonics2022, sahooHybridQuantumNanophotonic2023}, which inherits the advantages of its constituents, is constructed by means of pick and place transfer for evanescent coupling \cite{schellScanningProbebasedPickandplace2011, fehlerPurcellenhancedEmissionIndividual2020}, as well as for intrinsic coupling, positioning the quantum emitter in the field maximum of the photonic device \cite{fehlerHybridQuantumPhotonics2021}. Furthermore, the integration yield has been increased towards large numbers of individually addressable quantum emitters by lithographic positioning \cite{schrinnerIntegrationDiamondBasedQuantum2020} or designer nanodiamonds \cite{nganQuantumPhotonicCircuits2023}. However, until today access to stationary qubits was not demonstrated in coherent hybrid quantum photonics. \\
Here, we propose an \ac{nd}-hosted \siv's electron spin as a stationary qubit and experimentally demonstrate its access in hybrid quantum photonics entailing a number of advantages.
\siv in \nds are accompanied with a potential increase in spin coherence through modification of the phonon density of states \cite{klotzProlongedOrbitalRelaxation2022} or through strain \cite{sohnControllingCoherenceDiamond2018}. One dimensional \acp{pcc} enable coherent and efficient exchange of quantum information between stationary qubits and flying qubits in the framework of \cqed \cite{ritterElementaryQuantumNetwork2012}, while offering scalability and a small technological footprint on chip. \\
Our photonic base material is \SiN, which offers scalable fabrication and access to many key elements for quantum nodes integrated on a single chip \cite{blumenthalSiliconNitrideSilicon2018}. The one dimensional \acp{pcc} employed herein are formed in a single mode \SiN waveguide on top of SiO\textsubscript{2} with periodic elliptical holes patterned into the waveguide. These periodic holes are interleaved by a hole defect in the center of the cavity, leading to localized modes within the structure and serving as target region for \nd placement. At this center point, another single mode \SiN waveguide crosses the cavity waveguide orthogonally, being optimized for off-resonant excitation of the \sivs. Each of the two waveguides are equipped with efficient broadband out-of-plane couplers \cite{gehringBroadbandOutofplaneCoupling2019} to interface the nanostructure with Gaussian optics. For further details on the device \SuMa.
%
\begin{figure}
\includegraphics[scale=1]{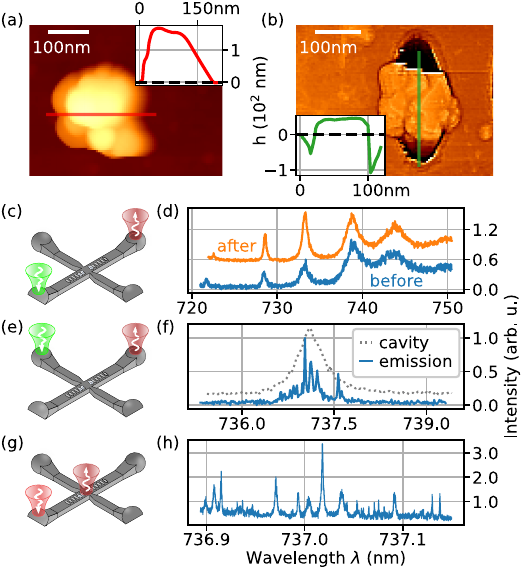}%
\caption{(a)~\afm scan and height profile (inset) of the \siv hosting \nds on a fused silica substrate. (b)~Parts of the same \nd embedded inside the first hole from the \pcc's center. The data is filtered with an unsharpen mask and the green line indicates the height profile path along the unfiltered data (inset). (c)~Measurement scheme to probe the cavity resonances using broadband \SiN fluorescence modulated by the cavity. (d)~Cavity spectrum before (blue) and after (orange, offset by 0.5) the pick and place procedure. (e)~Measurement scheme to detect cavity channeled \siv fluorescence via off-resonant excitation over the crossed waveguide. (f)~Normalized cavity channeled \siv florescence (blue) with the cavity tuned into resonance (dashed gray). (g)~Resonant excitation scheme with the laser coupled into the cavity via a coupler and fluoresence detection at the \pcc's center within the \siv phonon sideband. (h)~At cryogenic temperatures, laser scanning reveals distinct peaks signaling resonance with successfully placed \sivs.\label{fig:Placement}}
\end{figure}

Start of the assembly of our hybrid emitter-cavity system is identifying suitable \siv hosting \nds at room temperature via confocal microscopy. The procedure and \nd synthesis is described in the Supplemental Material \cite{SeeSupplementalMaterial}. Suitable candidates are transferred to the \pcc by \afm pick and place, following the method in reference \cite{fehlerHybridQuantumPhotonics2021}. In order to overcome limitations of evanescent coupling, the \nd of choice [\cref{fig:Placement} (a)] is placed inside a centered \pcc hole [\cref{fig:Placement} (b)]. After transfer the \nd is well embedded, as verified by its \afm height cross section [\cref{fig:Placement} (b), inset]. The whole \nd assembly does not alter the cavity properties, which we probe utilizing the \SiN inherent background fluorescence and the broadband out-of-plane couplers. A green laser is directed onto one of the cavity couplers and excites \SiN background fluorescence which is modulated by the cavity when collected at the opposing coupler [\cref{fig:Placement} (c)]. Quality factors $Q$ of the resonance modes, blue detuned from \SI{737}{\nm}, remain at $Q\geq1000$ while the resonance position is nearly unchanged after the assembly [\cref{fig:Placement} (d)]. Further details \SuMa.

\begin{figure*}
\includegraphics[scale=1]{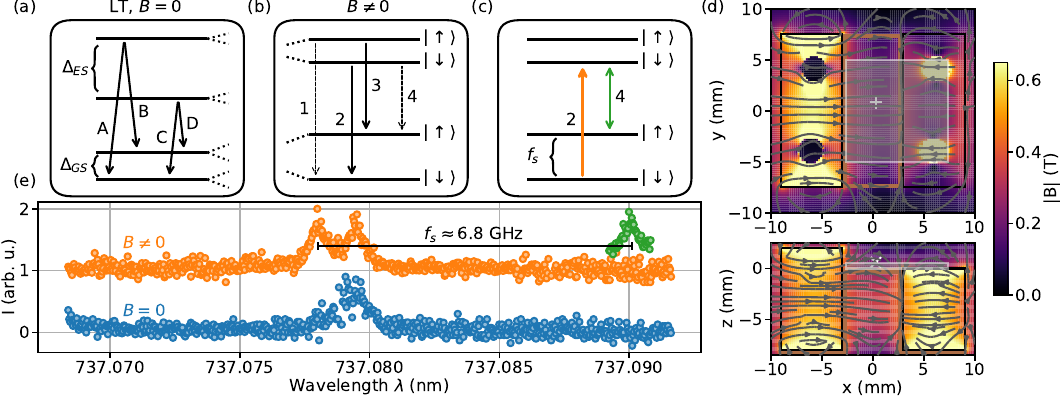}%
\caption{(a)~Level scheme of an \siv at cryogenic temperatures with four optical transitions, labeled A to D, connecting spin degenerate spin-orbit eigenstates separated by the strain dependent ground- ($\Delta_{\mathrm{GS}}$) and excited-state splitting~($\Delta_{\mathrm{ES}}$). (b)~In a magnetic field the level's the spin degeneracy is lifted resulting in four transitions (1-4) per level and enables optical access to the spin degree of freedom. (c)~Measurement scheme to probe the spin flipping transition 4. Continuous pumping on transition 2 (orange arrow) and probing for transition 4 (green arrow) with a second laser enables to determine the splitting $f_s$ of the spin-states. (d)~Simulation of the magnetic field from two permanent magnets (black rectangles) which are screwed to the sides of the coldfinger (copper colored lines), onto which the photonic chip (grey square) is mounted. White crosses mark the position of the \pcc. Simulations were performed using \textsc{magpylib} \cite{ortnerMagpylibFreePython2020}. (e)~Optical \siv transitions seen in a cavity \ple scan (blue) can split up when a magnetic field is applied (orange) due to a difference in spin splitting in the ground- and excited state. Pumping one of the spin-preserving transitions, and probing with a second laser reveals the spin-flipping transition (green). Data is normalized and offset. \label{fig:Magnetics}}
\end{figure*}

Successful placement of \sivs is verified at cryogenic temperatures ($\approx \SI{4}{\K}$) by off-resonant excitation via the pump waveguide and fluorescence collection at one of the cavity couplers [\cref{fig:Placement}~(e)] revealing cavity channeled fluorescence [\cref{fig:Placement}~(f)]. Alternatively a scanable titanium-sapphire laser is coupled into the cavity and fluorescence is detected from the \siv phonon sideband at the cavity center in a \ple experiment [\cref{fig:Placement} (g)]. An exemplary measurement is seen in \cref{fig:Placement} (h) revealing multiple peaks, corresponding to the laser hitting resonance with strain shifted \sivs in the \nd. Strain alters the \siv's ground- ($\Delta_{\mathrm{GS}}$) and excited state ($\Delta_{\mathrm{ES}}$) splitting [\cref{fig:Magnetics} (a)] as well as the weighted central wavelength (zero phonon line) of the optical transitions $A-D$ \cite{meesalaStrainEngineeringSiliconvacancy2018, rogersSingleSiVCenters2019}, which allows us to address individual \sivs.

The \siv's spin-orbit eigenstates, connected by these optical transitions, are spin degenerate in zero magnetic field \cite{heppElectronicStructureSilicon2014, mullerOpticalSignaturesSiliconvacancy2014, rogersAllOpticalInitializationReadout2014}. In a static magnetic field each eigenstate splits up due to the spin-1/2 nature of the \siv. The resulting spin-sublevels are denoted $\ket{\downarrow}$ and $\ket{\uparrow}$ herein [\cref{fig:Magnetics} (b)]. Transitions between these states are labeled $1-4$ in \cref{fig:Magnetics} (b) and are apparent for each of the optical transitions $A-D$. Here, the magnetic field is generated by a permanent neodymium magnet assembly, with simulated field strength above \SI{250}{\m\tesla} [\cref{fig:Magnetics} (d)]. The largest field component lies along the $x$ axis at the location of the \pcc \bSuMa{} which ensures that the magnetic field is parallel to well coupled \sivs with their dipole axis parallel to the \TE cavity mode. Alignment of the magnetic field to the \siv symmetry axis improves parameters, such as the spin-state purity \cite{heppElectronicStructureSilicon2014} and the spin relaxation time which influences the cyclicity of the optical transitions \cite{pingaultAllOpticalFormationCoherent2014}. Transition $2$ and $3$  potentially posses long cyclicity and connect levels of equal spin projection, hence also referred to as \textit{spin-preserving} transitions, while for transition $1$ and $4$ the spin will be flipped (\textit{spin-flipping}). The spin splitting in the ground- and excited state depends on the spin- as well as the orbital degree of freedom, where the latter is larger in the excited state \cite{heppElectronicStructureSilicon2014}. Consequently, transition 2 and 3 are spectrally separated enabling all-optical electron spin access \cite{rogersAllOpticalInitializationReadout2014}.
The intensity ratio of spin-flipping and spin-preserving transitions depends on the magnetic field alignment relative to the symmetry axis of the \siv, with the spin-flipping transition intensity being minimized for parallel alignment \cite{rogersAllOpticalInitializationReadout2014}. In this case [\cref{fig:Magnetics} (e)] one can probe and reveal the spin-flipping transition by continuously driving one of the spin-preserving transitions while sweeping a second laser's frequency [\cref{fig:Magnetics}~(c)]. The respective frequency difference yields the spin splitting $f_s \approx \SI{6.8}{\GHz}$ which is in agreement with the magnetic field simulations of \cref{fig:Magnetics}~(d).

\begin{figure}
\includegraphics[scale=1]{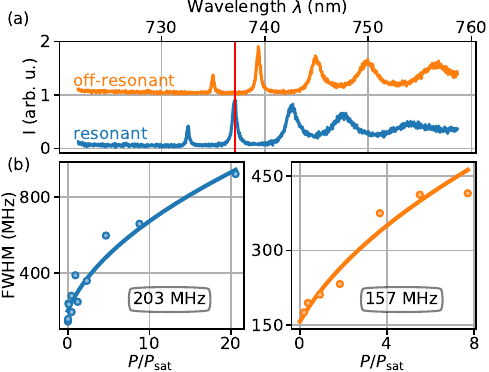}%
\caption{(a)~Cavity spectrum on- (blue) and off-resonance (orange) with the \siv. The vertical red line coincides with the corresponding transition wavelength. (b)~Power dependent \ple linewidth of the two spin-preserving transitions on- (blue) and off-resonant (orange) with the cavity for the \siv discussed in \cref{fig:Magnetics}. \label{fig:LineBroadening}}
\end{figure}
To determine the relevant \cqed parameter set \{$g$,~$\kappa$,~$\gamma$\} the linewidth for both spin-preserving transitions is measured at varying probe intensities on- and off-resonant with the cavity resonance [\cref{fig:LineBroadening}]. Extrapolating to zero power according to $\gamma = \gamma_0\sqrt{1+P/P_{\mathrm{sat}}}$ \cite{siegmanLasers1986} mitigates any influence of power broadening. Tuning the cavity resonance frequency is realized by controlled freezing of $\mathrm{N}_2$ to the \pcc\cite{mosorScanningPhotonicCrystal2005}, shifting the cavity resonances to longer wavelengths. Extracting $\gamma$ from the extrapolated zero power value yields $\gamma/2\pi = \SI{157\pm25}{\MHz}$ at a cavity detuning of $\Delta \approx 5\kappa$ [\cref{fig:LineBroadening}, orange]. This value is consistent with the Fourier-Transform limit of the \siv in \nds, where lifetimes in the range from \SI{200}{\ps} to \SI{2}{\ns} had been observed at room temperature \cite{lindnerStronglyInhomogeneousDistribution2018}. The cavity decay rate $\kappa$ is determined by the Lorentzian line shape of the resonance resulting in $\kappa/2\pi = \SI{273\pm7}{\GHz}$. From the zero power extrapolated Purcell broadened linewidth $\gamma_{\mathrm{on}}/2\pi = \SI{203\pm23}{\MHz}$ [\cref{fig:LineBroadening}~(b), orange] measured with the cavity in resonance ($\Delta = \SI{0.042\pm0.007}{}\cdot\kappa$) [\cref{fig:LineBroadening}~(a), blue], we can infer the cooperativity $C$ \cite{janitzCavityQuantumElectrodynamics2020} to be $C = \SI{0.30\pm0.24}{} = \frac{4g^2}{\kappa\gamma}$ translating to a single-photon Rabi frequency $g/2\pi = \SI{1.8\pm1.3}{\GHz}$ for the spin-preserving transitions. The complete \cqed parameter set reads 
\begin{equation*}
\{g, \kappa, \gamma\}/2\pi = \{1.8, 273, 0.157\}~\SI{}{\GHz}\;.
\end{equation*}

\begin{figure*}
\includegraphics[scale=1]{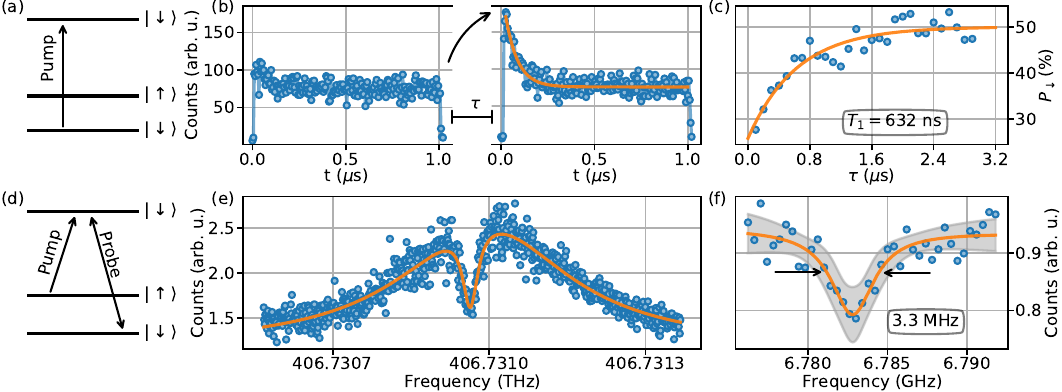}%
\caption{(a)~Measurement scheme to probe the spin relaxation time \Tone of the \siv. A laser resonant with transition 2 (see \cref{fig:Magnetics}) depopulates the $\ket{\downarrow}$ level and pumps population to $\ket{\uparrow}$. (b)~Time resolved fluorescence during the laser pulse. Increasing the pulse separation $\tau$ between consecutive laser pulses leads to increased fluorescence at the start of the laser pulse, due to exponential recovery towards thermal equilibrium during $\tau$. Fitting (orange, right hand panel) yields the spin initialization timescale of $\approx \SI{70}{\ns}$ (c)~Exponential recovery of population to thermal equilibrium versus laser pulse spacing $\tau$. The recovery timescale yields $T_1 = \SI{630\pm130}{\ns}$. (d)~\cpt experiment to extract the spin dephasing rate $\Gamma_2^*$. One laser fixed in resonance with e.g. transition 4 while another laser is scanned across the corresponding transition forming a $\Lambda$-type system. (e)~Sweeping the probe laser frequency yields the characteristic dip in the fluorescence signal as the system is trapped in the coherent superposition. (f)~Probing this dip width at low excitation powers and phase coherent \eom generated side bands reveals a minimum dip width of \SI{3.3\pm0.7}{\MHz}. Gray shaded area is the fit's $3\sigma$ confidence band. \label{fig:SpinAccess}}
\end{figure*}
All-optical spin access is realized by optical pumping on one of the spin-preserving transitions [\cref{fig:SpinAccess}~(a)], starting from equal population in $\ket{\downarrow}$ and $\ket{\uparrow}$ in thermal equilibrium at \SI{4}{\K} and a spin splitting of $f_s \approx \SI{6.8}{\GHz}$. During a \SI{1}{\us}-long laser pulse [\cref{fig:SpinAccess}~(b), right], this thermal equilibrium is disturbed by laser-induced optical spin pumping via the spin-flipping transition, depopulating the level $\ket{\downarrow}$. The fluorescence, collected in the phonon sideband, exponentially decays on a timescale of $\approx \SI{70}{\ns}$, corresponding to the spin initialization time. The steady state reached towards the end of the laser pulse yields a a final population in the spin-state $\ket{\uparrow}$ with a fidelity of $\approx \SI{75}{\percent}$, extracted from the ratio of steady state to peak signal. By applying consecutive laser pulses with a varying inter pulse delay, the characteristic relaxation time for the spin-state $T_1$ can be retrieved from an exponential fit of the corresponding initial peak heights [\cref{fig:SpinAccess}~(c)] which yields $T_1 = \SI{630\pm130}{\ns}$. $T_1$ is typically limited due to non-perfect alignment of the \siv symmetry axis and the magnetic field, which in our case indicates potential improvement in the coupling strength to the \pcc. As outlined earlier, ideally the magnetic field, the \siv symmetry axes and the \TE mode are co-linear. However, in the given configuration we attribute a decreased $T_1$ to a non-zero z-component ($\approx\SI{-10}{\degree}$) of the magnetic field \bSuMa. With vector control of the magnetic field, the misalignment could be determined and optimized by means of \afm nanomanipulation. In addition to the spin relaxation time $T_1$, the dephasing rate $\Gamma_2^*$ is of key interest, being the timescale dictating operations on the electronic spin. To probe $\Gamma_2^*$, a \acf{cpt} experiment is carried out where two laser fields are simultaneously applied to the \siv connecting a spin-preserving and spin-flipping transition in a $\Lambda$-type system [\cref{fig:SpinAccess}~(d)]. When both lasers fulfill the Raman-condition at zero detuning a coherent dark state is formed leading to a dip in fluorescence signal [\cref{fig:SpinAccess}~(e)]. The dip width is a direct measure of $\Gamma_2^*$  \cite{rogersAllOpticalInitializationReadout2014} when ensuring phase coherence of the laser fields and excluding power-broadening. This is realized by generating sidebands through microwave drive of an \eom and minimizing the excitation power. The dip width was measured to be as narrow as $\Gamma_2^* = \SI{3.3\pm0.7}{\MHz}$, which translates to $T_2^* = 1/(\pi \Gamma_2^*) = \SI{97\pm20}{\ns}$.

In conclusion, we demonstrate the fundamental step for an on-chip hybrid spin-photon interface based on the electronic spin of a \siv in \nds and \SiN photonics. We achieve a cooperativity of $C = \SI{0.30\pm0.24}{}$ and single-photon Rabi frequency of $g/2\pi = \SI{1.8\pm1.3}{\GHz}$ which both could be further improved by increasing the ratio of $Q$ factor to mode volume. The simultaneous requirements set herein for spin access and proper optical coupling to the \pcc restricts us to a small number of \sivs within the \nd. Using a higher density of \sivs or decreasing the crystal strain, which has direct influence on the possibility for optical spin access \cite{meesalaStrainEngineeringSiliconvacancy2018}, could increase the number of \siv candidates fulfilling both criteria. A further boost in coupling can be envisioned with additional alterations in the cavity design, or additional optimization through nanomanipulation steps \cite{hausslerPreparingSingleSiV2019} for dipole alignment. With the above turning nobs at hand, cooperativities well above one are within close reach. The accomplished spin initialization fidelity of $\approx \SI{75}{\percent}$ and the spin relaxation time $T_1$ of $\SI{630\pm130}{\ns}$ of our cavity coupled \siv could be improved by alignment of the magnetic field to the \siv symmetry axis through vector control. This should increase the spin relaxation time $T_1$ to orders of $\sim\SI{}{\ms}$ \cite{rogersAllOpticalInitializationReadout2014, metschInitializationReadoutNuclear2019} and enables long cyclicity of the spin-preserving transitions giving access to single-shot readout \cite{sukachevSiliconVacancySpinQubit2017}, while also increasing the upper bound for the spin coherence time $T_2$. The measured spin dephasing rate of $\Gamma_2^* = \SI{3.3\pm0.7}{\MHz}$ is in the range of values obtained for bulk diamond by optically detected magnetic resonance at $\sim\SI{2}{\K}$ in a bath cryostat \cite{metschInitializationReadoutNuclear2019}, suggesting the feasibility of integrating \sivs in \nds with potentially improved spin properties \cite{klotzProlongedOrbitalRelaxation2022}. Implementation of on-chip microwave structures to coherently control the electronic spin, or techniques relying on all optical driving \cite{debrouxQuantumControlTinVacancy2021, beckerAllOpticalControlSiliconVacancy2018}, together with above-mentioned  improvements leading to cooperativities well above one could enable spin-photon entangling gates based on spin-dependent cavity reflection signal \cite{nguyenIntegratedNanophotonicQuantum2019, nguyenQuantumNetworkNodes2019}.
\\
\begin{acknowledgments}
The project was funded by the BMBF/VDI in Project HybridQToken. A.K. acknowledges support of the  Baden-Wuerttemberg Stiftung gGmbH in Project No. BWST-ISF2018-008. The AFM was funded by the DFG, we thank Prof. Kay Gottschalk for the support. N.L. acknowledges support of the IQST. The authors thank V.A. Davydov for synthesis and processing of the nanodiamond material, and P. Maier for production of the FIB markers. D.W. acknowledges funding from the Deutsche Forschungsgemeinschaft (CRC 1459). H.G. acknowledges financial support of the Studienstiftung des deutschen Volkes. Experiments have been orchestrated by the open-source software Qudi \cite{binderQudiModularPython2017}.
\end{acknowledgments}
\nocite{fehlerPurcellenhancedEmissionIndividual2020, eichSinglePhotonEmissionIndividual2022, shiWiringPrecharacterizedSinglephoton2016}
\bibliography{AllOpticalHybridSpin.bib}

\begin{thebibliography}{39}%
\makeatletter
\providecommand \@ifxundefined [1]{%
 \@ifx{#1\undefined}
}%
\providecommand \@ifnum [1]{%
 \ifnum #1\expandafter \@firstoftwo
 \else \expandafter \@secondoftwo
 \fi
}%
\providecommand \@ifx [1]{%
 \ifx #1\expandafter \@firstoftwo
 \else \expandafter \@secondoftwo
 \fi
}%
\providecommand \natexlab [1]{#1}%
\providecommand \enquote  [1]{``#1''}%
\providecommand \bibnamefont  [1]{#1}%
\providecommand \bibfnamefont [1]{#1}%
\providecommand \citenamefont [1]{#1}%
\providecommand \href@noop [0]{\@secondoftwo}%
\providecommand \href [0]{\begingroup \@sanitize@url \@href}%
\providecommand \@href[1]{\@@startlink{#1}\@@href}%
\providecommand \@@href[1]{\endgroup#1\@@endlink}%
\providecommand \@sanitize@url [0]{\catcode `\\12\catcode `\$12\catcode
  `\&12\catcode `\#12\catcode `\^12\catcode `\_12\catcode `\%12\relax}%
\providecommand \@@startlink[1]{}%
\providecommand \@@endlink[0]{}%
\providecommand \url  [0]{\begingroup\@sanitize@url \@url }%
\providecommand \@url [1]{\endgroup\@href {#1}{\urlprefix }}%
\providecommand \urlprefix  [0]{URL }%
\providecommand \Eprint [0]{\href }%
\providecommand \doibase [0]{https://doi.org/}%
\providecommand \selectlanguage [0]{\@gobble}%
\providecommand \bibinfo  [0]{\@secondoftwo}%
\providecommand \bibfield  [0]{\@secondoftwo}%
\providecommand \translation [1]{[#1]}%
\providecommand \BibitemOpen [0]{}%
\providecommand \bibitemStop [0]{}%
\providecommand \bibitemNoStop [0]{.\EOS\space}%
\providecommand \EOS [0]{\spacefactor3000\relax}%
\providecommand \BibitemShut  [1]{\csname bibitem#1\endcsname}%
\let\auto@bib@innerbib\@empty
\bibitem [{\citenamefont {Kimble}(2008)}]{kimbleQuantumInternet2008}%
  \BibitemOpen
  \bibfield  {author} {\bibinfo {author} {\bibfnamefont {H.~J.}\ \bibnamefont
  {Kimble}},\ }\bibfield  {title} {\bibinfo {title} {The quantum internet},\
  }\href {https://doi.org/10.1038/nature07127} {\bibfield  {journal} {\bibinfo
  {journal} {Nature}\ }\textbf {\bibinfo {volume} {453}},\ \bibinfo {pages}
  {1023} (\bibinfo {year} {2008})}\BibitemShut {NoStop}%
\bibitem [{\citenamefont {Wehner}\ \emph {et~al.}(2018)\citenamefont {Wehner},
  \citenamefont {Elkouss},\ and\ \citenamefont
  {Hanson}}]{wehnerQuantumInternetVision2018}%
  \BibitemOpen
  \bibfield  {author} {\bibinfo {author} {\bibfnamefont {S.}~\bibnamefont
  {Wehner}}, \bibinfo {author} {\bibfnamefont {D.}~\bibnamefont {Elkouss}},\
  and\ \bibinfo {author} {\bibfnamefont {R.}~\bibnamefont {Hanson}},\
  }\bibfield  {title} {\bibinfo {title} {Quantum internet: {{A}} vision for the
  road ahead},\ }\href {https://doi.org/10.1126/science.aam9288} {\bibfield
  {journal} {\bibinfo  {journal} {Science}\ }\textbf {\bibinfo {volume}
  {362}},\ \bibinfo {pages} {eaam9288} (\bibinfo {year} {2018})}\BibitemShut
  {NoStop}%
\bibitem [{\citenamefont {Ruf}\ \emph {et~al.}(2021)\citenamefont {Ruf},
  \citenamefont {Wan}, \citenamefont {Choi}, \citenamefont {Englund},\ and\
  \citenamefont {Hanson}}]{rufQuantumNetworksBased2021}%
  \BibitemOpen
  \bibfield  {author} {\bibinfo {author} {\bibfnamefont {M.}~\bibnamefont
  {Ruf}}, \bibinfo {author} {\bibfnamefont {N.~H.}\ \bibnamefont {Wan}},
  \bibinfo {author} {\bibfnamefont {H.}~\bibnamefont {Choi}}, \bibinfo {author}
  {\bibfnamefont {D.}~\bibnamefont {Englund}},\ and\ \bibinfo {author}
  {\bibfnamefont {R.}~\bibnamefont {Hanson}},\ }\bibfield  {title} {\bibinfo
  {title} {Quantum networks based on color centers in diamond},\ }\href
  {https://doi.org/10.1063/5.0056534} {\bibfield  {journal} {\bibinfo
  {journal} {Journal of Applied Physics}\ }\textbf {\bibinfo {volume} {130}},\
  \bibinfo {pages} {070901} (\bibinfo {year} {2021})}\BibitemShut {NoStop}%
\bibitem [{\citenamefont {Northup}\ and\ \citenamefont
  {Blatt}(2014)}]{northupQuantumInformationTransfer2014}%
  \BibitemOpen
  \bibfield  {author} {\bibinfo {author} {\bibfnamefont {T.~E.}\ \bibnamefont
  {Northup}}\ and\ \bibinfo {author} {\bibfnamefont {R.}~\bibnamefont
  {Blatt}},\ }\bibfield  {title} {\bibinfo {title} {Quantum information
  transfer using photons},\ }\href {https://doi.org/10.1038/nphoton.2014.53}
  {\bibfield  {journal} {\bibinfo  {journal} {Nature Photon}\ }\textbf
  {\bibinfo {volume} {8}},\ \bibinfo {pages} {356} (\bibinfo {year}
  {2014})}\BibitemShut {NoStop}%
\bibitem [{\citenamefont {Rogers}\ \emph {et~al.}(2019)\citenamefont {Rogers},
  \citenamefont {Wang}, \citenamefont {Liu}, \citenamefont {Antoniuk},
  \citenamefont {Osterkamp}, \citenamefont {Davydov}, \citenamefont {Agafonov},
  \citenamefont {Filipovski}, \citenamefont {Jelezko},\ and\ \citenamefont
  {Kubanek}}]{rogersSingleSiVCenters2019}%
  \BibitemOpen
  \bibfield  {author} {\bibinfo {author} {\bibfnamefont {L.~J.}\ \bibnamefont
  {Rogers}}, \bibinfo {author} {\bibfnamefont {O.}~\bibnamefont {Wang}},
  \bibinfo {author} {\bibfnamefont {Y.}~\bibnamefont {Liu}}, \bibinfo {author}
  {\bibfnamefont {L.}~\bibnamefont {Antoniuk}}, \bibinfo {author}
  {\bibfnamefont {C.}~\bibnamefont {Osterkamp}}, \bibinfo {author}
  {\bibfnamefont {V.~A.}\ \bibnamefont {Davydov}}, \bibinfo {author}
  {\bibfnamefont {V.~N.}\ \bibnamefont {Agafonov}}, \bibinfo {author}
  {\bibfnamefont {A.~B.}\ \bibnamefont {Filipovski}}, \bibinfo {author}
  {\bibfnamefont {F.}~\bibnamefont {Jelezko}},\ and\ \bibinfo {author}
  {\bibfnamefont {A.}~\bibnamefont {Kubanek}},\ }\bibfield  {title} {\bibinfo
  {title} {Single {{SiV- Centers}} in {{Low-Strain Nanodiamonds}} with
  {{Bulklike Spectral Properties}} and {{Nanomanipulation Capabilities}}},\
  }\href {https://doi.org/10.1103/PhysRevApplied.11.024073} {\bibfield
  {journal} {\bibinfo  {journal} {Phys. Rev. Applied}\ }\textbf {\bibinfo
  {volume} {11}},\ \bibinfo {pages} {024073} (\bibinfo {year}
  {2019})}\BibitemShut {NoStop}%
\bibitem [{\citenamefont {Waltrich}\ \emph {et~al.}(2023)\citenamefont
  {Waltrich}, \citenamefont {Klotz}, \citenamefont {Agafonov},\ and\
  \citenamefont {Kubanek}}]{waltrichTwophotonInterferenceSiliconvacancy2023}%
  \BibitemOpen
  \bibfield  {author} {\bibinfo {author} {\bibfnamefont {R.}~\bibnamefont
  {Waltrich}}, \bibinfo {author} {\bibfnamefont {M.}~\bibnamefont {Klotz}},
  \bibinfo {author} {\bibfnamefont {V.~N.}\ \bibnamefont {Agafonov}},\ and\
  \bibinfo {author} {\bibfnamefont {A.}~\bibnamefont {Kubanek}},\ }\bibfield
  {title} {\bibinfo {title} {Two-photon interference from silicon-vacancy
  centers in remote nanodiamonds},\ }\bibfield  {journal} {\bibinfo  {journal}
  {Nanophotonics}\ }\href {https://doi.org/10.1515/nanoph-2023-0379}
  {10.1515/nanoph-2023-0379} (\bibinfo {year} {2023})\BibitemShut {NoStop}%
\bibitem [{\citenamefont {Kubanek}\ \emph {et~al.}(2022)\citenamefont
  {Kubanek}, \citenamefont {Ovvyan}, \citenamefont {Antoniuk}, \citenamefont
  {Lettner},\ and\ \citenamefont
  {Pernice}}]{kubanekHybridQuantumNanophotonics2022}%
  \BibitemOpen
  \bibfield  {author} {\bibinfo {author} {\bibfnamefont {A.}~\bibnamefont
  {Kubanek}}, \bibinfo {author} {\bibfnamefont {A.~P.}\ \bibnamefont {Ovvyan}},
  \bibinfo {author} {\bibfnamefont {L.}~\bibnamefont {Antoniuk}}, \bibinfo
  {author} {\bibfnamefont {N.}~\bibnamefont {Lettner}},\ and\ \bibinfo {author}
  {\bibfnamefont {W.~H.~P.}\ \bibnamefont {Pernice}},\ }\bibfield  {title}
  {\bibinfo {title} {Hybrid {{Quantum Nanophotonics}}\textemdash{{Interfacing
  Color Center}} in {{Nanodiamonds}} with {{Si3N4-Photonics}}},\ }in\ \href
  {https://doi.org/10.1007/978-3-031-16518-4_5} {\emph {\bibinfo {booktitle}
  {Progress in {{Nanophotonics}} 7}}},\ \bibinfo {series and number} {Topics in
  {{Applied Physics}}},\ \bibinfo {editor} {edited by\ \bibinfo {editor}
  {\bibfnamefont {T.}~\bibnamefont {Yatsui}}}\ (\bibinfo  {publisher}
  {{Springer International Publishing}},\ \bibinfo {address} {{Cham}},\
  \bibinfo {year} {2022})\ pp.\ \bibinfo {pages} {123--174}\BibitemShut
  {NoStop}%
\bibitem [{\citenamefont {Sahoo}\ \emph {et~al.}(2023)\citenamefont {Sahoo},
  \citenamefont {Davydov}, \citenamefont {Agafonov},\ and\ \citenamefont
  {Bogdanov}}]{sahooHybridQuantumNanophotonic2023}%
  \BibitemOpen
  \bibfield  {author} {\bibinfo {author} {\bibfnamefont {S.}~\bibnamefont
  {Sahoo}}, \bibinfo {author} {\bibfnamefont {V.~A.}\ \bibnamefont {Davydov}},
  \bibinfo {author} {\bibfnamefont {V.~N.}\ \bibnamefont {Agafonov}},\ and\
  \bibinfo {author} {\bibfnamefont {S.~I.}\ \bibnamefont {Bogdanov}},\
  }\bibfield  {title} {\bibinfo {title} {Hybrid quantum nanophotonic devices
  with color centers in nanodiamonds [{{Invited}}]},\ }\href
  {https://doi.org/10.1364/OME.471376} {\bibfield  {journal} {\bibinfo
  {journal} {Opt. Mater. Express, OME}\ }\textbf {\bibinfo {volume} {13}},\
  \bibinfo {pages} {191} (\bibinfo {year} {2023})}\BibitemShut {NoStop}%
\bibitem [{\citenamefont {Schell}\ \emph {et~al.}(2011)\citenamefont {Schell},
  \citenamefont {Kewes}, \citenamefont {Schr{\"o}der}, \citenamefont {Wolters},
  \citenamefont {Aichele},\ and\ \citenamefont
  {Benson}}]{schellScanningProbebasedPickandplace2011}%
  \BibitemOpen
  \bibfield  {author} {\bibinfo {author} {\bibfnamefont {A.~W.}\ \bibnamefont
  {Schell}}, \bibinfo {author} {\bibfnamefont {G.}~\bibnamefont {Kewes}},
  \bibinfo {author} {\bibfnamefont {T.}~\bibnamefont {Schr{\"o}der}}, \bibinfo
  {author} {\bibfnamefont {J.}~\bibnamefont {Wolters}}, \bibinfo {author}
  {\bibfnamefont {T.}~\bibnamefont {Aichele}},\ and\ \bibinfo {author}
  {\bibfnamefont {O.}~\bibnamefont {Benson}},\ }\bibfield  {title} {\bibinfo
  {title} {A scanning probe-based pick-and-place procedure for assembly of
  integrated quantum optical hybrid devices},\ }\href
  {https://doi.org/10.1063/1.3615629} {\bibfield  {journal} {\bibinfo
  {journal} {Review of Scientific Instruments}\ }\textbf {\bibinfo {volume}
  {82}},\ \bibinfo {pages} {073709} (\bibinfo {year} {2011})}\BibitemShut
  {NoStop}%
\bibitem [{\citenamefont {Fehler}\ \emph {et~al.}(2020)\citenamefont {Fehler},
  \citenamefont {Ovvyan}, \citenamefont {Antoniuk}, \citenamefont {Lettner},
  \citenamefont {Gruhler}, \citenamefont {Davydov}, \citenamefont {Agafonov},
  \citenamefont {Pernice},\ and\ \citenamefont
  {Kubanek}}]{fehlerPurcellenhancedEmissionIndividual2020}%
  \BibitemOpen
  \bibfield  {author} {\bibinfo {author} {\bibfnamefont {K.~G.}\ \bibnamefont
  {Fehler}}, \bibinfo {author} {\bibfnamefont {A.~P.}\ \bibnamefont {Ovvyan}},
  \bibinfo {author} {\bibfnamefont {L.}~\bibnamefont {Antoniuk}}, \bibinfo
  {author} {\bibfnamefont {N.}~\bibnamefont {Lettner}}, \bibinfo {author}
  {\bibfnamefont {N.}~\bibnamefont {Gruhler}}, \bibinfo {author} {\bibfnamefont
  {V.~A.}\ \bibnamefont {Davydov}}, \bibinfo {author} {\bibfnamefont {V.~N.}\
  \bibnamefont {Agafonov}}, \bibinfo {author} {\bibfnamefont {W.~H.~P.}\
  \bibnamefont {Pernice}},\ and\ \bibinfo {author} {\bibfnamefont
  {A.}~\bibnamefont {Kubanek}},\ }\bibfield  {title} {\bibinfo {title}
  {Purcell-enhanced emission from individual {{SiV}}- center in nanodiamonds
  coupled to a {{Si3N4-based}}, photonic crystal cavity},\ }\href
  {https://doi.org/10.1515/nanoph-2020-0257} {\bibfield  {journal} {\bibinfo
  {journal} {Nanophotonics}\ }\textbf {\bibinfo {volume} {9}},\ \bibinfo
  {pages} {3655} (\bibinfo {year} {2020})}\BibitemShut {NoStop}%
\bibitem [{\citenamefont {Fehler}\ \emph {et~al.}(2021)\citenamefont {Fehler},
  \citenamefont {Antoniuk}, \citenamefont {Lettner}, \citenamefont {Ovvyan},
  \citenamefont {Waltrich}, \citenamefont {Gruhler}, \citenamefont {Davydov},
  \citenamefont {Agafonov}, \citenamefont {Pernice},\ and\ \citenamefont
  {Kubanek}}]{fehlerHybridQuantumPhotonics2021}%
  \BibitemOpen
  \bibfield  {author} {\bibinfo {author} {\bibfnamefont {K.~G.}\ \bibnamefont
  {Fehler}}, \bibinfo {author} {\bibfnamefont {L.}~\bibnamefont {Antoniuk}},
  \bibinfo {author} {\bibfnamefont {N.}~\bibnamefont {Lettner}}, \bibinfo
  {author} {\bibfnamefont {A.~P.}\ \bibnamefont {Ovvyan}}, \bibinfo {author}
  {\bibfnamefont {R.}~\bibnamefont {Waltrich}}, \bibinfo {author}
  {\bibfnamefont {N.}~\bibnamefont {Gruhler}}, \bibinfo {author} {\bibfnamefont
  {V.~A.}\ \bibnamefont {Davydov}}, \bibinfo {author} {\bibfnamefont {V.~N.}\
  \bibnamefont {Agafonov}}, \bibinfo {author} {\bibfnamefont {W.~H.~P.}\
  \bibnamefont {Pernice}},\ and\ \bibinfo {author} {\bibfnamefont
  {A.}~\bibnamefont {Kubanek}},\ }\bibfield  {title} {\bibinfo {title} {Hybrid
  {{Quantum Photonics Based}} on {{Artificial Atoms Placed Inside One Hole}} of
  a {{Photonic Crystal Cavity}}},\ }\href
  {https://doi.org/10.1021/acsphotonics.1c00530} {\bibfield  {journal}
  {\bibinfo  {journal} {ACS Photonics}\ }\textbf {\bibinfo {volume} {8}},\
  \bibinfo {pages} {2635} (\bibinfo {year} {2021})}\BibitemShut {NoStop}%
\bibitem [{\citenamefont {Schrinner}\ \emph {et~al.}(2020)\citenamefont
  {Schrinner}, \citenamefont {Olthaus}, \citenamefont {Reiter},\ and\
  \citenamefont {Schuck}}]{schrinnerIntegrationDiamondBasedQuantum2020}%
  \BibitemOpen
  \bibfield  {author} {\bibinfo {author} {\bibfnamefont {P.~P.~J.}\
  \bibnamefont {Schrinner}}, \bibinfo {author} {\bibfnamefont {J.}~\bibnamefont
  {Olthaus}}, \bibinfo {author} {\bibfnamefont {D.~E.}\ \bibnamefont
  {Reiter}},\ and\ \bibinfo {author} {\bibfnamefont {C.}~\bibnamefont
  {Schuck}},\ }\bibfield  {title} {\bibinfo {title} {Integration of
  {{Diamond-Based Quantum Emitters}} with {{Nanophotonic Circuits}}},\ }\href
  {https://doi.org/10.1021/acs.nanolett.0c03262} {\bibfield  {journal}
  {\bibinfo  {journal} {Nano Lett.}\ }\textbf {\bibinfo {volume} {20}},\
  \bibinfo {pages} {8170} (\bibinfo {year} {2020})}\BibitemShut {NoStop}%
\bibitem [{\citenamefont {Ngan}\ \emph {et~al.}(2023)\citenamefont {Ngan},
  \citenamefont {Zhan}, \citenamefont {Dory}, \citenamefont {Vu{\v
  c}kovi{\'c}},\ and\ \citenamefont {Sun}}]{nganQuantumPhotonicCircuits2023}%
  \BibitemOpen
  \bibfield  {author} {\bibinfo {author} {\bibfnamefont {K.}~\bibnamefont
  {Ngan}}, \bibinfo {author} {\bibfnamefont {Y.}~\bibnamefont {Zhan}}, \bibinfo
  {author} {\bibfnamefont {C.}~\bibnamefont {Dory}}, \bibinfo {author}
  {\bibfnamefont {J.}~\bibnamefont {Vu{\v c}kovi{\'c}}},\ and\ \bibinfo
  {author} {\bibfnamefont {S.}~\bibnamefont {Sun}},\ }\href
  {https://doi.org/10.48550/arXiv.2307.13309} {\bibinfo {title} {Quantum
  {{Photonic Circuits Integrated}} with {{Color Centers}} in {{Designer
  Nanodiamonds}}}} (\bibinfo {year} {2023}),\ \Eprint
  {https://arxiv.org/abs/2307.13309} {arxiv:2307.13309 [physics,
  physics:quant-ph]} \BibitemShut {NoStop}%
\bibitem [{\citenamefont {Klotz}\ \emph {et~al.}(2022)\citenamefont {Klotz},
  \citenamefont {Fehler}, \citenamefont {Waltrich}, \citenamefont {Steiger},
  \citenamefont {H{\"a}u{\ss}ler}, \citenamefont {Reddy}, \citenamefont
  {Kulikova}, \citenamefont {Davydov}, \citenamefont {Agafonov}, \citenamefont
  {Doherty},\ and\ \citenamefont
  {Kubanek}}]{klotzProlongedOrbitalRelaxation2022}%
  \BibitemOpen
  \bibfield  {author} {\bibinfo {author} {\bibfnamefont {M.}~\bibnamefont
  {Klotz}}, \bibinfo {author} {\bibfnamefont {K.~G.}\ \bibnamefont {Fehler}},
  \bibinfo {author} {\bibfnamefont {R.}~\bibnamefont {Waltrich}}, \bibinfo
  {author} {\bibfnamefont {E.~S.}\ \bibnamefont {Steiger}}, \bibinfo {author}
  {\bibfnamefont {S.}~\bibnamefont {H{\"a}u{\ss}ler}}, \bibinfo {author}
  {\bibfnamefont {P.}~\bibnamefont {Reddy}}, \bibinfo {author} {\bibfnamefont
  {L.~F.}\ \bibnamefont {Kulikova}}, \bibinfo {author} {\bibfnamefont {V.~A.}\
  \bibnamefont {Davydov}}, \bibinfo {author} {\bibfnamefont {V.~N.}\
  \bibnamefont {Agafonov}}, \bibinfo {author} {\bibfnamefont {M.~W.}\
  \bibnamefont {Doherty}},\ and\ \bibinfo {author} {\bibfnamefont
  {A.}~\bibnamefont {Kubanek}},\ }\bibfield  {title} {\bibinfo {title}
  {Prolonged {{Orbital Relaxation}} by {{Locally Modified Phonon Density}} of
  {{States}} for the {{SiV- Center}} in {{Nanodiamonds}}},\ }\href
  {https://doi.org/10.1103/PhysRevLett.128.153602} {\bibfield  {journal}
  {\bibinfo  {journal} {Phys. Rev. Lett.}\ }\textbf {\bibinfo {volume} {128}},\
  \bibinfo {pages} {153602} (\bibinfo {year} {2022})}\BibitemShut {NoStop}%
\bibitem [{\citenamefont {Sohn}\ \emph {et~al.}(2018)\citenamefont {Sohn},
  \citenamefont {Meesala}, \citenamefont {Pingault}, \citenamefont {Atikian},
  \citenamefont {Holzgrafe}, \citenamefont {G{\"u}ndo{\u g}an}, \citenamefont
  {Stavrakas}, \citenamefont {Stanley}, \citenamefont {Sipahigil},
  \citenamefont {Choi}, \citenamefont {Zhang}, \citenamefont {Pacheco},
  \citenamefont {Abraham}, \citenamefont {Bielejec}, \citenamefont {Lukin},
  \citenamefont {Atat{\"u}re},\ and\ \citenamefont {Lon{\v
  c}ar}}]{sohnControllingCoherenceDiamond2018}%
  \BibitemOpen
  \bibfield  {author} {\bibinfo {author} {\bibfnamefont {Y.-I.}\ \bibnamefont
  {Sohn}}, \bibinfo {author} {\bibfnamefont {S.}~\bibnamefont {Meesala}},
  \bibinfo {author} {\bibfnamefont {B.}~\bibnamefont {Pingault}}, \bibinfo
  {author} {\bibfnamefont {H.~A.}\ \bibnamefont {Atikian}}, \bibinfo {author}
  {\bibfnamefont {J.}~\bibnamefont {Holzgrafe}}, \bibinfo {author}
  {\bibfnamefont {M.}~\bibnamefont {G{\"u}ndo{\u g}an}}, \bibinfo {author}
  {\bibfnamefont {C.}~\bibnamefont {Stavrakas}}, \bibinfo {author}
  {\bibfnamefont {M.~J.}\ \bibnamefont {Stanley}}, \bibinfo {author}
  {\bibfnamefont {A.}~\bibnamefont {Sipahigil}}, \bibinfo {author}
  {\bibfnamefont {J.}~\bibnamefont {Choi}}, \bibinfo {author} {\bibfnamefont
  {M.}~\bibnamefont {Zhang}}, \bibinfo {author} {\bibfnamefont {J.~L.}\
  \bibnamefont {Pacheco}}, \bibinfo {author} {\bibfnamefont {J.}~\bibnamefont
  {Abraham}}, \bibinfo {author} {\bibfnamefont {E.}~\bibnamefont {Bielejec}},
  \bibinfo {author} {\bibfnamefont {M.~D.}\ \bibnamefont {Lukin}}, \bibinfo
  {author} {\bibfnamefont {M.}~\bibnamefont {Atat{\"u}re}},\ and\ \bibinfo
  {author} {\bibfnamefont {M.}~\bibnamefont {Lon{\v c}ar}},\ }\bibfield
  {title} {\bibinfo {title} {Controlling the coherence of a diamond spin qubit
  through its strain environment},\ }\href
  {https://doi.org/10.1038/s41467-018-04340-3} {\bibfield  {journal} {\bibinfo
  {journal} {Nat Commun}\ }\textbf {\bibinfo {volume} {9}},\ \bibinfo {pages}
  {1} (\bibinfo {year} {2018})}\BibitemShut {NoStop}%
\bibitem [{\citenamefont {Ritter}\ \emph {et~al.}(2012)\citenamefont {Ritter},
  \citenamefont {N{\"o}lleke}, \citenamefont {Hahn}, \citenamefont {Reiserer},
  \citenamefont {Neuzner}, \citenamefont {Uphoff}, \citenamefont {M{\"u}cke},
  \citenamefont {Figueroa}, \citenamefont {Bochmann},\ and\ \citenamefont
  {Rempe}}]{ritterElementaryQuantumNetwork2012}%
  \BibitemOpen
  \bibfield  {author} {\bibinfo {author} {\bibfnamefont {S.}~\bibnamefont
  {Ritter}}, \bibinfo {author} {\bibfnamefont {C.}~\bibnamefont {N{\"o}lleke}},
  \bibinfo {author} {\bibfnamefont {C.}~\bibnamefont {Hahn}}, \bibinfo {author}
  {\bibfnamefont {A.}~\bibnamefont {Reiserer}}, \bibinfo {author}
  {\bibfnamefont {A.}~\bibnamefont {Neuzner}}, \bibinfo {author} {\bibfnamefont
  {M.}~\bibnamefont {Uphoff}}, \bibinfo {author} {\bibfnamefont
  {M.}~\bibnamefont {M{\"u}cke}}, \bibinfo {author} {\bibfnamefont
  {E.}~\bibnamefont {Figueroa}}, \bibinfo {author} {\bibfnamefont
  {J.}~\bibnamefont {Bochmann}},\ and\ \bibinfo {author} {\bibfnamefont
  {G.}~\bibnamefont {Rempe}},\ }\bibfield  {title} {\bibinfo {title} {An
  elementary quantum network of single atoms in optical cavities},\ }\href
  {https://doi.org/10.1038/nature11023} {\bibfield  {journal} {\bibinfo
  {journal} {Nature}\ }\textbf {\bibinfo {volume} {484}},\ \bibinfo {pages}
  {195} (\bibinfo {year} {2012})}\BibitemShut {NoStop}%
\bibitem [{\citenamefont {Blumenthal}\ \emph {et~al.}(2018)\citenamefont
  {Blumenthal}, \citenamefont {Heideman}, \citenamefont {Geuzebroek},
  \citenamefont {Leinse},\ and\ \citenamefont
  {Roeloffzen}}]{blumenthalSiliconNitrideSilicon2018}%
  \BibitemOpen
  \bibfield  {author} {\bibinfo {author} {\bibfnamefont {D.~J.}\ \bibnamefont
  {Blumenthal}}, \bibinfo {author} {\bibfnamefont {R.}~\bibnamefont
  {Heideman}}, \bibinfo {author} {\bibfnamefont {D.}~\bibnamefont
  {Geuzebroek}}, \bibinfo {author} {\bibfnamefont {A.}~\bibnamefont {Leinse}},\
  and\ \bibinfo {author} {\bibfnamefont {C.}~\bibnamefont {Roeloffzen}},\
  }\bibfield  {title} {\bibinfo {title} {Silicon {{Nitride}} in {{Silicon
  Photonics}}},\ }\href {https://doi.org/10.1109/JPROC.2018.2861576} {\bibfield
   {journal} {\bibinfo  {journal} {Proceedings of the IEEE}\ }\textbf {\bibinfo
  {volume} {106}},\ \bibinfo {pages} {2209} (\bibinfo {year}
  {2018})}\BibitemShut {NoStop}%
\bibitem [{\citenamefont {Gehring}\ \emph {et~al.}(2019)\citenamefont
  {Gehring}, \citenamefont {Eich}, \citenamefont {Schuck},\ and\ \citenamefont
  {Pernice}}]{gehringBroadbandOutofplaneCoupling2019}%
  \BibitemOpen
  \bibfield  {author} {\bibinfo {author} {\bibfnamefont {H.}~\bibnamefont
  {Gehring}}, \bibinfo {author} {\bibfnamefont {A.}~\bibnamefont {Eich}},
  \bibinfo {author} {\bibfnamefont {C.}~\bibnamefont {Schuck}},\ and\ \bibinfo
  {author} {\bibfnamefont {W.~H.~P.}\ \bibnamefont {Pernice}},\ }\bibfield
  {title} {\bibinfo {title} {Broadband out-of-plane coupling at visible
  wavelengths},\ }\href {https://doi.org/10.1364/OL.44.005089} {\bibfield
  {journal} {\bibinfo  {journal} {Opt. Lett., OL}\ }\textbf {\bibinfo {volume}
  {44}},\ \bibinfo {pages} {5089} (\bibinfo {year} {2019})}\BibitemShut
  {NoStop}%
\bibitem [{See()}]{SeeSupplementalMaterial}%
  \BibitemOpen
  \href@noop {} {\bibinfo {title} {See {{Supplemental Material}} at [{{URL}}
  will be inserted by publisher] for details on the setup, device assembly,
  nanodiamond and sample fabrication and magentic field components, which
  includes {{Refs}}. [38-39]}}\BibitemShut {NoStop}%
\bibitem [{\citenamefont {Meesala}\ \emph {et~al.}(2018)\citenamefont
  {Meesala}, \citenamefont {Sohn}, \citenamefont {Pingault}, \citenamefont
  {Shao}, \citenamefont {Atikian}, \citenamefont {Holzgrafe}, \citenamefont
  {G{\"u}ndo{\u g}an}, \citenamefont {Stavrakas}, \citenamefont {Sipahigil},
  \citenamefont {Chia}, \citenamefont {Evans}, \citenamefont {Burek},
  \citenamefont {Zhang}, \citenamefont {Wu}, \citenamefont {Pacheco},
  \citenamefont {Abraham}, \citenamefont {Bielejec}, \citenamefont {Lukin},
  \citenamefont {Atat{\"u}re},\ and\ \citenamefont {Lon{\v
  c}ar}}]{meesalaStrainEngineeringSiliconvacancy2018}%
  \BibitemOpen
  \bibfield  {author} {\bibinfo {author} {\bibfnamefont {S.}~\bibnamefont
  {Meesala}}, \bibinfo {author} {\bibfnamefont {Y.-I.}\ \bibnamefont {Sohn}},
  \bibinfo {author} {\bibfnamefont {B.}~\bibnamefont {Pingault}}, \bibinfo
  {author} {\bibfnamefont {L.}~\bibnamefont {Shao}}, \bibinfo {author}
  {\bibfnamefont {H.~A.}\ \bibnamefont {Atikian}}, \bibinfo {author}
  {\bibfnamefont {J.}~\bibnamefont {Holzgrafe}}, \bibinfo {author}
  {\bibfnamefont {M.}~\bibnamefont {G{\"u}ndo{\u g}an}}, \bibinfo {author}
  {\bibfnamefont {C.}~\bibnamefont {Stavrakas}}, \bibinfo {author}
  {\bibfnamefont {A.}~\bibnamefont {Sipahigil}}, \bibinfo {author}
  {\bibfnamefont {C.}~\bibnamefont {Chia}}, \bibinfo {author} {\bibfnamefont
  {R.}~\bibnamefont {Evans}}, \bibinfo {author} {\bibfnamefont {M.~J.}\
  \bibnamefont {Burek}}, \bibinfo {author} {\bibfnamefont {M.}~\bibnamefont
  {Zhang}}, \bibinfo {author} {\bibfnamefont {L.}~\bibnamefont {Wu}}, \bibinfo
  {author} {\bibfnamefont {J.~L.}\ \bibnamefont {Pacheco}}, \bibinfo {author}
  {\bibfnamefont {J.}~\bibnamefont {Abraham}}, \bibinfo {author} {\bibfnamefont
  {E.}~\bibnamefont {Bielejec}}, \bibinfo {author} {\bibfnamefont {M.~D.}\
  \bibnamefont {Lukin}}, \bibinfo {author} {\bibfnamefont {M.}~\bibnamefont
  {Atat{\"u}re}},\ and\ \bibinfo {author} {\bibfnamefont {M.}~\bibnamefont
  {Lon{\v c}ar}},\ }\bibfield  {title} {\bibinfo {title} {Strain engineering of
  the silicon-vacancy center in diamond},\ }\href
  {https://doi.org/10.1103/PhysRevB.97.205444} {\bibfield  {journal} {\bibinfo
  {journal} {Phys. Rev. B}\ }\textbf {\bibinfo {volume} {97}},\ \bibinfo
  {pages} {205444} (\bibinfo {year} {2018})}\BibitemShut {NoStop}%
\bibitem [{\citenamefont {Ortner}\ and\ \citenamefont
  {Coliado~Bandeira}(2020)}]{ortnerMagpylibFreePython2020}%
  \BibitemOpen
  \bibfield  {author} {\bibinfo {author} {\bibfnamefont {M.}~\bibnamefont
  {Ortner}}\ and\ \bibinfo {author} {\bibfnamefont {L.~G.}\ \bibnamefont
  {Coliado~Bandeira}},\ }\bibfield  {title} {\bibinfo {title} {Magpylib: {{A}}
  free {{Python}} package for magnetic field computation},\ }\href
  {https://doi.org/10.1016/j.softx.2020.100466} {\bibfield  {journal} {\bibinfo
   {journal} {SoftwareX}\ }\textbf {\bibinfo {volume} {11}},\ \bibinfo {pages}
  {100466} (\bibinfo {year} {2020})}\BibitemShut {NoStop}%
\bibitem [{\citenamefont {Hepp}\ \emph {et~al.}(2014)\citenamefont {Hepp},
  \citenamefont {M{\"u}ller}, \citenamefont {Waselowski}, \citenamefont
  {Becker}, \citenamefont {Pingault}, \citenamefont {Sternschulte},
  \citenamefont {{Steinm{\"u}ller-Nethl}}, \citenamefont {Gali}, \citenamefont
  {Maze}, \citenamefont {Atat{\"u}re},\ and\ \citenamefont
  {Becher}}]{heppElectronicStructureSilicon2014}%
  \BibitemOpen
  \bibfield  {author} {\bibinfo {author} {\bibfnamefont {C.}~\bibnamefont
  {Hepp}}, \bibinfo {author} {\bibfnamefont {T.}~\bibnamefont {M{\"u}ller}},
  \bibinfo {author} {\bibfnamefont {V.}~\bibnamefont {Waselowski}}, \bibinfo
  {author} {\bibfnamefont {J.~N.}\ \bibnamefont {Becker}}, \bibinfo {author}
  {\bibfnamefont {B.}~\bibnamefont {Pingault}}, \bibinfo {author}
  {\bibfnamefont {H.}~\bibnamefont {Sternschulte}}, \bibinfo {author}
  {\bibfnamefont {D.}~\bibnamefont {{Steinm{\"u}ller-Nethl}}}, \bibinfo
  {author} {\bibfnamefont {A.}~\bibnamefont {Gali}}, \bibinfo {author}
  {\bibfnamefont {J.~R.}\ \bibnamefont {Maze}}, \bibinfo {author}
  {\bibfnamefont {M.}~\bibnamefont {Atat{\"u}re}},\ and\ \bibinfo {author}
  {\bibfnamefont {C.}~\bibnamefont {Becher}},\ }\bibfield  {title} {\bibinfo
  {title} {Electronic {{Structure}} of the {{Silicon Vacancy Color Center}} in
  {{Diamond}}},\ }\bibfield  {journal} {\bibinfo  {journal} {Physical Review
  Letters}\ }\textbf {\bibinfo {volume} {112}},\ \href
  {https://doi.org/10.1103/PhysRevLett.112.036405}
  {10.1103/PhysRevLett.112.036405} (\bibinfo {year} {2014})\BibitemShut
  {NoStop}%
\bibitem [{\citenamefont {M{\"u}ller}\ \emph {et~al.}(2014)\citenamefont
  {M{\"u}ller}, \citenamefont {Hepp}, \citenamefont {Pingault}, \citenamefont
  {Neu}, \citenamefont {Gsell}, \citenamefont {Schreck}, \citenamefont
  {Sternschulte}, \citenamefont {{Steinm{\"u}ller-Nethl}}, \citenamefont
  {Becher},\ and\ \citenamefont
  {Atat{\"u}re}}]{mullerOpticalSignaturesSiliconvacancy2014}%
  \BibitemOpen
  \bibfield  {author} {\bibinfo {author} {\bibfnamefont {T.}~\bibnamefont
  {M{\"u}ller}}, \bibinfo {author} {\bibfnamefont {C.}~\bibnamefont {Hepp}},
  \bibinfo {author} {\bibfnamefont {B.}~\bibnamefont {Pingault}}, \bibinfo
  {author} {\bibfnamefont {E.}~\bibnamefont {Neu}}, \bibinfo {author}
  {\bibfnamefont {S.}~\bibnamefont {Gsell}}, \bibinfo {author} {\bibfnamefont
  {M.}~\bibnamefont {Schreck}}, \bibinfo {author} {\bibfnamefont
  {H.}~\bibnamefont {Sternschulte}}, \bibinfo {author} {\bibfnamefont
  {D.}~\bibnamefont {{Steinm{\"u}ller-Nethl}}}, \bibinfo {author}
  {\bibfnamefont {C.}~\bibnamefont {Becher}},\ and\ \bibinfo {author}
  {\bibfnamefont {M.}~\bibnamefont {Atat{\"u}re}},\ }\bibfield  {title}
  {\bibinfo {title} {Optical signatures of silicon-vacancy spins in diamond},\
  }\href {https://doi.org/10.1038/ncomms4328} {\bibfield  {journal} {\bibinfo
  {journal} {Nat Commun}\ }\textbf {\bibinfo {volume} {5}},\ \bibinfo {pages}
  {3328} (\bibinfo {year} {2014})}\BibitemShut {NoStop}%
\bibitem [{\citenamefont {Rogers}\ \emph {et~al.}(2014)\citenamefont {Rogers},
  \citenamefont {Jahnke}, \citenamefont {Metsch}, \citenamefont {Sipahigil},
  \citenamefont {Binder}, \citenamefont {Teraji}, \citenamefont {Sumiya},
  \citenamefont {Isoya}, \citenamefont {Lukin}, \citenamefont {Hemmer},\ and\
  \citenamefont {Jelezko}}]{rogersAllOpticalInitializationReadout2014}%
  \BibitemOpen
  \bibfield  {author} {\bibinfo {author} {\bibfnamefont {L.~J.}\ \bibnamefont
  {Rogers}}, \bibinfo {author} {\bibfnamefont {K.~D.}\ \bibnamefont {Jahnke}},
  \bibinfo {author} {\bibfnamefont {M.~H.}\ \bibnamefont {Metsch}}, \bibinfo
  {author} {\bibfnamefont {A.}~\bibnamefont {Sipahigil}}, \bibinfo {author}
  {\bibfnamefont {J.~M.}\ \bibnamefont {Binder}}, \bibinfo {author}
  {\bibfnamefont {T.}~\bibnamefont {Teraji}}, \bibinfo {author} {\bibfnamefont
  {H.}~\bibnamefont {Sumiya}}, \bibinfo {author} {\bibfnamefont
  {J.}~\bibnamefont {Isoya}}, \bibinfo {author} {\bibfnamefont {M.~D.}\
  \bibnamefont {Lukin}}, \bibinfo {author} {\bibfnamefont {P.}~\bibnamefont
  {Hemmer}},\ and\ \bibinfo {author} {\bibfnamefont {F.}~\bibnamefont
  {Jelezko}},\ }\bibfield  {title} {\bibinfo {title} {All-{{Optical
  Initialization}}, {{Readout}}, and {{Coherent Preparation}} of {{Single
  Silicon-Vacancy Spins}} in {{Diamond}}},\ }\href
  {https://doi.org/10.1103/PhysRevLett.113.263602} {\bibfield  {journal}
  {\bibinfo  {journal} {Phys. Rev. Lett.}\ }\textbf {\bibinfo {volume} {113}},\
  \bibinfo {pages} {263602} (\bibinfo {year} {2014})}\BibitemShut {NoStop}%
\bibitem [{\citenamefont {Pingault}\ \emph {et~al.}(2014)\citenamefont
  {Pingault}, \citenamefont {Becker}, \citenamefont {Schulte}, \citenamefont
  {Arend}, \citenamefont {Hepp}, \citenamefont {Godde}, \citenamefont
  {Tartakovskii}, \citenamefont {Markham}, \citenamefont {Becher},\ and\
  \citenamefont {Atat{\"u}re}}]{pingaultAllOpticalFormationCoherent2014}%
  \BibitemOpen
  \bibfield  {author} {\bibinfo {author} {\bibfnamefont {B.}~\bibnamefont
  {Pingault}}, \bibinfo {author} {\bibfnamefont {J.~N.}\ \bibnamefont
  {Becker}}, \bibinfo {author} {\bibfnamefont {C.~H.~H.}\ \bibnamefont
  {Schulte}}, \bibinfo {author} {\bibfnamefont {C.}~\bibnamefont {Arend}},
  \bibinfo {author} {\bibfnamefont {C.}~\bibnamefont {Hepp}}, \bibinfo {author}
  {\bibfnamefont {T.}~\bibnamefont {Godde}}, \bibinfo {author} {\bibfnamefont
  {A.~I.}\ \bibnamefont {Tartakovskii}}, \bibinfo {author} {\bibfnamefont
  {M.}~\bibnamefont {Markham}}, \bibinfo {author} {\bibfnamefont
  {C.}~\bibnamefont {Becher}},\ and\ \bibinfo {author} {\bibfnamefont
  {M.}~\bibnamefont {Atat{\"u}re}},\ }\bibfield  {title} {\bibinfo {title}
  {All-{{Optical Formation}} of {{Coherent Dark States}} of {{Silicon-Vacancy
  Spins}} in {{Diamond}}},\ }\href
  {https://doi.org/10.1103/PhysRevLett.113.263601} {\bibfield  {journal}
  {\bibinfo  {journal} {Phys. Rev. Lett.}\ }\textbf {\bibinfo {volume} {113}},\
  \bibinfo {pages} {263601} (\bibinfo {year} {2014})}\BibitemShut {NoStop}%
\bibitem [{\citenamefont {Siegman}(1986)}]{siegmanLasers1986}%
  \BibitemOpen
  \bibfield  {author} {\bibinfo {author} {\bibfnamefont {A.~E.}\ \bibnamefont
  {Siegman}},\ }\href@noop {} {\emph {\bibinfo {title} {Lasers}}}\ (\bibinfo
  {publisher} {{University Science Books}},\ \bibinfo {year}
  {1986})\BibitemShut {NoStop}%
\bibitem [{\citenamefont {Mosor}\ \emph {et~al.}(2005)\citenamefont {Mosor},
  \citenamefont {Hendrickson}, \citenamefont {Richards}, \citenamefont {Sweet},
  \citenamefont {Khitrova}, \citenamefont {Gibbs}, \citenamefont {Yoshie},
  \citenamefont {Scherer}, \citenamefont {Shchekin},\ and\ \citenamefont
  {Deppe}}]{mosorScanningPhotonicCrystal2005}%
  \BibitemOpen
  \bibfield  {author} {\bibinfo {author} {\bibfnamefont {S.}~\bibnamefont
  {Mosor}}, \bibinfo {author} {\bibfnamefont {J.}~\bibnamefont {Hendrickson}},
  \bibinfo {author} {\bibfnamefont {B.~C.}\ \bibnamefont {Richards}}, \bibinfo
  {author} {\bibfnamefont {J.}~\bibnamefont {Sweet}}, \bibinfo {author}
  {\bibfnamefont {G.}~\bibnamefont {Khitrova}}, \bibinfo {author}
  {\bibfnamefont {H.~M.}\ \bibnamefont {Gibbs}}, \bibinfo {author}
  {\bibfnamefont {T.}~\bibnamefont {Yoshie}}, \bibinfo {author} {\bibfnamefont
  {A.}~\bibnamefont {Scherer}}, \bibinfo {author} {\bibfnamefont {O.~B.}\
  \bibnamefont {Shchekin}},\ and\ \bibinfo {author} {\bibfnamefont {D.~G.}\
  \bibnamefont {Deppe}},\ }\bibfield  {title} {\bibinfo {title} {Scanning a
  photonic crystal slab nanocavity by condensation of xenon},\ }\href
  {https://doi.org/10.1063/1.2076435} {\bibfield  {journal} {\bibinfo
  {journal} {Appl. Phys. Lett.}\ }\textbf {\bibinfo {volume} {87}},\ \bibinfo
  {pages} {141105} (\bibinfo {year} {2005})}\BibitemShut {NoStop}%
\bibitem [{\citenamefont {Lindner}\ \emph {et~al.}(2018)\citenamefont
  {Lindner}, \citenamefont {Bommer}, \citenamefont {Muzha}, \citenamefont
  {Krueger}, \citenamefont {Gines}, \citenamefont {Mandal}, \citenamefont
  {Williams}, \citenamefont {Londero}, \citenamefont {Gali},\ and\
  \citenamefont {Becher}}]{lindnerStronglyInhomogeneousDistribution2018}%
  \BibitemOpen
  \bibfield  {author} {\bibinfo {author} {\bibfnamefont {S.}~\bibnamefont
  {Lindner}}, \bibinfo {author} {\bibfnamefont {A.}~\bibnamefont {Bommer}},
  \bibinfo {author} {\bibfnamefont {A.}~\bibnamefont {Muzha}}, \bibinfo
  {author} {\bibfnamefont {A.}~\bibnamefont {Krueger}}, \bibinfo {author}
  {\bibfnamefont {L.}~\bibnamefont {Gines}}, \bibinfo {author} {\bibfnamefont
  {S.}~\bibnamefont {Mandal}}, \bibinfo {author} {\bibfnamefont
  {O.}~\bibnamefont {Williams}}, \bibinfo {author} {\bibfnamefont
  {E.}~\bibnamefont {Londero}}, \bibinfo {author} {\bibfnamefont
  {A.}~\bibnamefont {Gali}},\ and\ \bibinfo {author} {\bibfnamefont
  {C.}~\bibnamefont {Becher}},\ }\bibfield  {title} {\bibinfo {title} {Strongly
  inhomogeneous distribution of spectral properties of silicon-vacancy color
  centers in nanodiamonds},\ }\href {https://doi.org/10.1088/1367-2630/aae93f}
  {\bibfield  {journal} {\bibinfo  {journal} {New J. Phys.}\ }\textbf {\bibinfo
  {volume} {20}},\ \bibinfo {pages} {115002} (\bibinfo {year}
  {2018})}\BibitemShut {NoStop}%
\bibitem [{\citenamefont {Janitz}\ \emph {et~al.}(2020)\citenamefont {Janitz},
  \citenamefont {Bhaskar},\ and\ \citenamefont
  {Childress}}]{janitzCavityQuantumElectrodynamics2020}%
  \BibitemOpen
  \bibfield  {author} {\bibinfo {author} {\bibfnamefont {E.}~\bibnamefont
  {Janitz}}, \bibinfo {author} {\bibfnamefont {M.~K.}\ \bibnamefont
  {Bhaskar}},\ and\ \bibinfo {author} {\bibfnamefont {L.}~\bibnamefont
  {Childress}},\ }\bibfield  {title} {\bibinfo {title} {Cavity quantum
  electrodynamics with color centers in diamond},\ }\href
  {https://doi.org/10.1364/OPTICA.398628} {\bibfield  {journal} {\bibinfo
  {journal} {Optica}\ }\textbf {\bibinfo {volume} {7}},\ \bibinfo {pages}
  {1232} (\bibinfo {year} {2020})}\BibitemShut {NoStop}%
\bibitem [{\citenamefont {H{\"a}u{\ss}ler}\ \emph {et~al.}(2019)\citenamefont
  {H{\"a}u{\ss}ler}, \citenamefont {Hartung}, \citenamefont {Fehler},
  \citenamefont {Antoniuk}, \citenamefont {Kulikova}, \citenamefont {Davydov},
  \citenamefont {Agafonov}, \citenamefont {Jelezko},\ and\ \citenamefont
  {Kubanek}}]{hausslerPreparingSingleSiV2019}%
  \BibitemOpen
  \bibfield  {author} {\bibinfo {author} {\bibfnamefont {S.}~\bibnamefont
  {H{\"a}u{\ss}ler}}, \bibinfo {author} {\bibfnamefont {L.}~\bibnamefont
  {Hartung}}, \bibinfo {author} {\bibfnamefont {K.~G.}\ \bibnamefont {Fehler}},
  \bibinfo {author} {\bibfnamefont {L.}~\bibnamefont {Antoniuk}}, \bibinfo
  {author} {\bibfnamefont {L.~F.}\ \bibnamefont {Kulikova}}, \bibinfo {author}
  {\bibfnamefont {V.~A.}\ \bibnamefont {Davydov}}, \bibinfo {author}
  {\bibfnamefont {V.~N.}\ \bibnamefont {Agafonov}}, \bibinfo {author}
  {\bibfnamefont {F.}~\bibnamefont {Jelezko}},\ and\ \bibinfo {author}
  {\bibfnamefont {A.}~\bibnamefont {Kubanek}},\ }\bibfield  {title} {\bibinfo
  {title} {Preparing single {{SiV}}- center in nanodiamonds for external,
  optical coupling with access to all degrees of freedom},\ }\href
  {https://doi.org/10.1088/1367-2630/ab4cf7} {\bibfield  {journal} {\bibinfo
  {journal} {New J. Phys.}\ }\textbf {\bibinfo {volume} {21}},\ \bibinfo
  {pages} {103047} (\bibinfo {year} {2019})}\BibitemShut {NoStop}%
\bibitem [{\citenamefont {Metsch}\ \emph {et~al.}(2019)\citenamefont {Metsch},
  \citenamefont {Senkalla}, \citenamefont {Tratzmiller}, \citenamefont
  {Scheuer}, \citenamefont {Kern}, \citenamefont {Achard}, \citenamefont
  {Tallaire}, \citenamefont {Plenio}, \citenamefont {Siyushev},\ and\
  \citenamefont {Jelezko}}]{metschInitializationReadoutNuclear2019}%
  \BibitemOpen
  \bibfield  {author} {\bibinfo {author} {\bibfnamefont {M.~H.}\ \bibnamefont
  {Metsch}}, \bibinfo {author} {\bibfnamefont {K.}~\bibnamefont {Senkalla}},
  \bibinfo {author} {\bibfnamefont {B.}~\bibnamefont {Tratzmiller}}, \bibinfo
  {author} {\bibfnamefont {J.}~\bibnamefont {Scheuer}}, \bibinfo {author}
  {\bibfnamefont {M.}~\bibnamefont {Kern}}, \bibinfo {author} {\bibfnamefont
  {J.}~\bibnamefont {Achard}}, \bibinfo {author} {\bibfnamefont
  {A.}~\bibnamefont {Tallaire}}, \bibinfo {author} {\bibfnamefont {M.~B.}\
  \bibnamefont {Plenio}}, \bibinfo {author} {\bibfnamefont {P.}~\bibnamefont
  {Siyushev}},\ and\ \bibinfo {author} {\bibfnamefont {F.}~\bibnamefont
  {Jelezko}},\ }\bibfield  {title} {\bibinfo {title} {Initialization and
  {{Readout}} of {{Nuclear Spins}} via a {{Negatively Charged Silicon-Vacancy
  Center}} in {{Diamond}}},\ }\href
  {https://doi.org/10.1103/PhysRevLett.122.190503} {\bibfield  {journal}
  {\bibinfo  {journal} {Phys. Rev. Lett.}\ }\textbf {\bibinfo {volume} {122}},\
  \bibinfo {pages} {190503} (\bibinfo {year} {2019})}\BibitemShut {NoStop}%
\bibitem [{\citenamefont {Sukachev}\ \emph {et~al.}(2017)\citenamefont
  {Sukachev}, \citenamefont {Sipahigil}, \citenamefont {Nguyen}, \citenamefont
  {Bhaskar}, \citenamefont {Evans}, \citenamefont {Jelezko},\ and\
  \citenamefont {Lukin}}]{sukachevSiliconVacancySpinQubit2017}%
  \BibitemOpen
  \bibfield  {author} {\bibinfo {author} {\bibfnamefont {D.~D.}\ \bibnamefont
  {Sukachev}}, \bibinfo {author} {\bibfnamefont {A.}~\bibnamefont {Sipahigil}},
  \bibinfo {author} {\bibfnamefont {C.~T.}\ \bibnamefont {Nguyen}}, \bibinfo
  {author} {\bibfnamefont {M.~K.}\ \bibnamefont {Bhaskar}}, \bibinfo {author}
  {\bibfnamefont {R.~E.}\ \bibnamefont {Evans}}, \bibinfo {author}
  {\bibfnamefont {F.}~\bibnamefont {Jelezko}},\ and\ \bibinfo {author}
  {\bibfnamefont {M.~D.}\ \bibnamefont {Lukin}},\ }\bibfield  {title} {\bibinfo
  {title} {Silicon-{{Vacancy Spin Qubit}} in {{Diamond}}: {{A Quantum Memory
  Exceeding}} 10 ms with {{Single-Shot State Readout}}},\ }\href
  {https://doi.org/10.1103/PhysRevLett.119.223602} {\bibfield  {journal}
  {\bibinfo  {journal} {Phys. Rev. Lett.}\ }\textbf {\bibinfo {volume} {119}},\
  \bibinfo {pages} {223602} (\bibinfo {year} {2017})}\BibitemShut {NoStop}%
\bibitem [{\citenamefont {Debroux}\ \emph {et~al.}(2021)\citenamefont
  {Debroux}, \citenamefont {Michaels}, \citenamefont {Purser}, \citenamefont
  {Wan}, \citenamefont {Trusheim}, \citenamefont {Arjona~Mart{\'i}nez},
  \citenamefont {Parker}, \citenamefont {Stramma}, \citenamefont {Chen},
  \citenamefont {{de Santis}}, \citenamefont {Alexeev}, \citenamefont
  {Ferrari}, \citenamefont {Englund}, \citenamefont {Gangloff},\ and\
  \citenamefont {Atat{\"u}re}}]{debrouxQuantumControlTinVacancy2021}%
  \BibitemOpen
  \bibfield  {author} {\bibinfo {author} {\bibfnamefont {R.}~\bibnamefont
  {Debroux}}, \bibinfo {author} {\bibfnamefont {C.~P.}\ \bibnamefont
  {Michaels}}, \bibinfo {author} {\bibfnamefont {C.~M.}\ \bibnamefont
  {Purser}}, \bibinfo {author} {\bibfnamefont {N.}~\bibnamefont {Wan}},
  \bibinfo {author} {\bibfnamefont {M.~E.}\ \bibnamefont {Trusheim}}, \bibinfo
  {author} {\bibfnamefont {J.}~\bibnamefont {Arjona~Mart{\'i}nez}}, \bibinfo
  {author} {\bibfnamefont {R.~A.}\ \bibnamefont {Parker}}, \bibinfo {author}
  {\bibfnamefont {A.~M.}\ \bibnamefont {Stramma}}, \bibinfo {author}
  {\bibfnamefont {K.~C.}\ \bibnamefont {Chen}}, \bibinfo {author}
  {\bibfnamefont {L.}~\bibnamefont {{de Santis}}}, \bibinfo {author}
  {\bibfnamefont {E.~M.}\ \bibnamefont {Alexeev}}, \bibinfo {author}
  {\bibfnamefont {A.~C.}\ \bibnamefont {Ferrari}}, \bibinfo {author}
  {\bibfnamefont {D.}~\bibnamefont {Englund}}, \bibinfo {author} {\bibfnamefont
  {D.~A.}\ \bibnamefont {Gangloff}},\ and\ \bibinfo {author} {\bibfnamefont
  {M.}~\bibnamefont {Atat{\"u}re}},\ }\bibfield  {title} {\bibinfo {title}
  {Quantum {{Control}} of the {{Tin-Vacancy Spin Qubit}} in {{Diamond}}},\
  }\href {https://doi.org/10.1103/PhysRevX.11.041041} {\bibfield  {journal}
  {\bibinfo  {journal} {Phys. Rev. X}\ }\textbf {\bibinfo {volume} {11}},\
  \bibinfo {pages} {041041} (\bibinfo {year} {2021})}\BibitemShut {NoStop}%
\bibitem [{\citenamefont {Becker}\ \emph {et~al.}(2018)\citenamefont {Becker},
  \citenamefont {Pingault}, \citenamefont {Gro{\ss}}, \citenamefont
  {G{\"u}ndo{\u g}an}, \citenamefont {Kukharchyk}, \citenamefont {Markham},
  \citenamefont {Edmonds}, \citenamefont {Atat{\"u}re}, \citenamefont
  {Bushev},\ and\ \citenamefont
  {Becher}}]{beckerAllOpticalControlSiliconVacancy2018}%
  \BibitemOpen
  \bibfield  {author} {\bibinfo {author} {\bibfnamefont {J.~N.}\ \bibnamefont
  {Becker}}, \bibinfo {author} {\bibfnamefont {B.}~\bibnamefont {Pingault}},
  \bibinfo {author} {\bibfnamefont {D.}~\bibnamefont {Gro{\ss}}}, \bibinfo
  {author} {\bibfnamefont {M.}~\bibnamefont {G{\"u}ndo{\u g}an}}, \bibinfo
  {author} {\bibfnamefont {N.}~\bibnamefont {Kukharchyk}}, \bibinfo {author}
  {\bibfnamefont {M.}~\bibnamefont {Markham}}, \bibinfo {author} {\bibfnamefont
  {A.}~\bibnamefont {Edmonds}}, \bibinfo {author} {\bibfnamefont
  {M.}~\bibnamefont {Atat{\"u}re}}, \bibinfo {author} {\bibfnamefont
  {P.}~\bibnamefont {Bushev}},\ and\ \bibinfo {author} {\bibfnamefont
  {C.}~\bibnamefont {Becher}},\ }\bibfield  {title} {\bibinfo {title}
  {All-{{Optical Control}} of the {{Silicon-Vacancy Spin}} in {{Diamond}} at
  {{Millikelvin Temperatures}}},\ }\href
  {https://doi.org/10.1103/PhysRevLett.120.053603} {\bibfield  {journal}
  {\bibinfo  {journal} {Phys. Rev. Lett.}\ }\textbf {\bibinfo {volume} {120}},\
  \bibinfo {pages} {053603} (\bibinfo {year} {2018})}\BibitemShut {NoStop}%
\bibitem [{\citenamefont {Nguyen}\ \emph
  {et~al.}(2019{\natexlab{a}})\citenamefont {Nguyen}, \citenamefont {Sukachev},
  \citenamefont {Bhaskar}, \citenamefont {Machielse}, \citenamefont {Levonian},
  \citenamefont {Knall}, \citenamefont {Stroganov}, \citenamefont {Chia},
  \citenamefont {Burek}, \citenamefont {Riedinger}, \citenamefont {Park},
  \citenamefont {Lon{\v c}ar},\ and\ \citenamefont
  {Lukin}}]{nguyenIntegratedNanophotonicQuantum2019}%
  \BibitemOpen
  \bibfield  {author} {\bibinfo {author} {\bibfnamefont {C.~T.}\ \bibnamefont
  {Nguyen}}, \bibinfo {author} {\bibfnamefont {D.~D.}\ \bibnamefont
  {Sukachev}}, \bibinfo {author} {\bibfnamefont {M.~K.}\ \bibnamefont
  {Bhaskar}}, \bibinfo {author} {\bibfnamefont {B.}~\bibnamefont {Machielse}},
  \bibinfo {author} {\bibfnamefont {D.~S.}\ \bibnamefont {Levonian}}, \bibinfo
  {author} {\bibfnamefont {E.~N.}\ \bibnamefont {Knall}}, \bibinfo {author}
  {\bibfnamefont {P.}~\bibnamefont {Stroganov}}, \bibinfo {author}
  {\bibfnamefont {C.}~\bibnamefont {Chia}}, \bibinfo {author} {\bibfnamefont
  {M.~J.}\ \bibnamefont {Burek}}, \bibinfo {author} {\bibfnamefont
  {R.}~\bibnamefont {Riedinger}}, \bibinfo {author} {\bibfnamefont
  {H.}~\bibnamefont {Park}}, \bibinfo {author} {\bibfnamefont {M.}~\bibnamefont
  {Lon{\v c}ar}},\ and\ \bibinfo {author} {\bibfnamefont {M.~D.}\ \bibnamefont
  {Lukin}},\ }\bibfield  {title} {\bibinfo {title} {An integrated nanophotonic
  quantum register based on silicon-vacancy spins in diamond},\ }\href
  {https://doi.org/10.1103/PhysRevB.100.165428} {\bibfield  {journal} {\bibinfo
   {journal} {Phys. Rev. B}\ }\textbf {\bibinfo {volume} {100}},\ \bibinfo
  {pages} {165428} (\bibinfo {year} {2019}{\natexlab{a}})}\BibitemShut
  {NoStop}%
\bibitem [{\citenamefont {Nguyen}\ \emph
  {et~al.}(2019{\natexlab{b}})\citenamefont {Nguyen}, \citenamefont {Sukachev},
  \citenamefont {Bhaskar}, \citenamefont {Machielse}, \citenamefont {Levonian},
  \citenamefont {Knall}, \citenamefont {Stroganov}, \citenamefont {Riedinger},
  \citenamefont {Park}, \citenamefont {Lon{\v c}ar},\ and\ \citenamefont
  {Lukin}}]{nguyenQuantumNetworkNodes2019}%
  \BibitemOpen
  \bibfield  {author} {\bibinfo {author} {\bibfnamefont {C.~T.}\ \bibnamefont
  {Nguyen}}, \bibinfo {author} {\bibfnamefont {D.~D.}\ \bibnamefont
  {Sukachev}}, \bibinfo {author} {\bibfnamefont {M.~K.}\ \bibnamefont
  {Bhaskar}}, \bibinfo {author} {\bibfnamefont {B.}~\bibnamefont {Machielse}},
  \bibinfo {author} {\bibfnamefont {D.~S.}\ \bibnamefont {Levonian}}, \bibinfo
  {author} {\bibfnamefont {E.~N.}\ \bibnamefont {Knall}}, \bibinfo {author}
  {\bibfnamefont {P.}~\bibnamefont {Stroganov}}, \bibinfo {author}
  {\bibfnamefont {R.}~\bibnamefont {Riedinger}}, \bibinfo {author}
  {\bibfnamefont {H.}~\bibnamefont {Park}}, \bibinfo {author} {\bibfnamefont
  {M.}~\bibnamefont {Lon{\v c}ar}},\ and\ \bibinfo {author} {\bibfnamefont
  {M.~D.}\ \bibnamefont {Lukin}},\ }\bibfield  {title} {\bibinfo {title}
  {Quantum {{Network Nodes Based}} on {{Diamond Qubits}} with an {{Efficient
  Nanophotonic Interface}}},\ }\href
  {https://doi.org/10.1103/PhysRevLett.123.183602} {\bibfield  {journal}
  {\bibinfo  {journal} {Phys. Rev. Lett.}\ }\textbf {\bibinfo {volume} {123}},\
  \bibinfo {pages} {183602} (\bibinfo {year} {2019}{\natexlab{b}})}\BibitemShut
  {NoStop}%
\bibitem [{\citenamefont {Binder}\ \emph {et~al.}(2017)\citenamefont {Binder},
  \citenamefont {Stark}, \citenamefont {Tomek}, \citenamefont {Scheuer},
  \citenamefont {Frank}, \citenamefont {Jahnke}, \citenamefont {M{\"u}ller},
  \citenamefont {Schmitt}, \citenamefont {Metsch}, \citenamefont {Unden},
  \citenamefont {Gehring}, \citenamefont {Huck}, \citenamefont {Andersen},
  \citenamefont {Rogers},\ and\ \citenamefont
  {Jelezko}}]{binderQudiModularPython2017}%
  \BibitemOpen
  \bibfield  {author} {\bibinfo {author} {\bibfnamefont {J.~M.}\ \bibnamefont
  {Binder}}, \bibinfo {author} {\bibfnamefont {A.}~\bibnamefont {Stark}},
  \bibinfo {author} {\bibfnamefont {N.}~\bibnamefont {Tomek}}, \bibinfo
  {author} {\bibfnamefont {J.}~\bibnamefont {Scheuer}}, \bibinfo {author}
  {\bibfnamefont {F.}~\bibnamefont {Frank}}, \bibinfo {author} {\bibfnamefont
  {K.~D.}\ \bibnamefont {Jahnke}}, \bibinfo {author} {\bibfnamefont
  {C.}~\bibnamefont {M{\"u}ller}}, \bibinfo {author} {\bibfnamefont
  {S.}~\bibnamefont {Schmitt}}, \bibinfo {author} {\bibfnamefont {M.~H.}\
  \bibnamefont {Metsch}}, \bibinfo {author} {\bibfnamefont {T.}~\bibnamefont
  {Unden}}, \bibinfo {author} {\bibfnamefont {T.}~\bibnamefont {Gehring}},
  \bibinfo {author} {\bibfnamefont {A.}~\bibnamefont {Huck}}, \bibinfo {author}
  {\bibfnamefont {U.~L.}\ \bibnamefont {Andersen}}, \bibinfo {author}
  {\bibfnamefont {L.~J.}\ \bibnamefont {Rogers}},\ and\ \bibinfo {author}
  {\bibfnamefont {F.}~\bibnamefont {Jelezko}},\ }\bibfield  {title} {\bibinfo
  {title} {Qudi: {{A}} modular python suite for experiment control and data
  processing},\ }\href {https://doi.org/10.1016/j.softx.2017.02.001} {\bibfield
   {journal} {\bibinfo  {journal} {SoftwareX}\ }\textbf {\bibinfo {volume}
  {6}},\ \bibinfo {pages} {85} (\bibinfo {year} {2017})}\BibitemShut {NoStop}%
\bibitem [{\citenamefont {Eich}\ \emph {et~al.}(2022)\citenamefont {Eich},
  \citenamefont {Spiekermann}, \citenamefont {Gehring}, \citenamefont {Sommer},
  \citenamefont {Bankwitz}, \citenamefont {Schrinner}, \citenamefont
  {Preu{\ss}}, \citenamefont {{Michaelis de Vasconcellos}}, \citenamefont
  {Bratschitsch}, \citenamefont {Pernice},\ and\ \citenamefont
  {Schuck}}]{eichSinglePhotonEmissionIndividual2022}%
  \BibitemOpen
  \bibfield  {author} {\bibinfo {author} {\bibfnamefont {A.}~\bibnamefont
  {Eich}}, \bibinfo {author} {\bibfnamefont {T.~C.}\ \bibnamefont
  {Spiekermann}}, \bibinfo {author} {\bibfnamefont {H.}~\bibnamefont
  {Gehring}}, \bibinfo {author} {\bibfnamefont {L.}~\bibnamefont {Sommer}},
  \bibinfo {author} {\bibfnamefont {J.~R.}\ \bibnamefont {Bankwitz}}, \bibinfo
  {author} {\bibfnamefont {P.~P.~J.}\ \bibnamefont {Schrinner}}, \bibinfo
  {author} {\bibfnamefont {J.~A.}\ \bibnamefont {Preu{\ss}}}, \bibinfo {author}
  {\bibfnamefont {S.}~\bibnamefont {{Michaelis de Vasconcellos}}}, \bibinfo
  {author} {\bibfnamefont {R.}~\bibnamefont {Bratschitsch}}, \bibinfo {author}
  {\bibfnamefont {W.~H.~P.}\ \bibnamefont {Pernice}},\ and\ \bibinfo {author}
  {\bibfnamefont {C.}~\bibnamefont {Schuck}},\ }\bibfield  {title} {\bibinfo
  {title} {Single-{{Photon Emission}} from {{Individual Nanophotonic-Integrated
  Colloidal Quantum Dots}}},\ }\href
  {https://doi.org/10.1021/acsphotonics.1c01493} {\bibfield  {journal}
  {\bibinfo  {journal} {ACS Photonics}\ }\textbf {\bibinfo {volume} {9}},\
  \bibinfo {pages} {551} (\bibinfo {year} {2022})}\BibitemShut {NoStop}%
\bibitem [{\citenamefont {Shi}\ \emph {et~al.}(2016)\citenamefont {Shi},
  \citenamefont {Sontheimer}, \citenamefont {Nikolay}, \citenamefont {Schell},
  \citenamefont {Fischer}, \citenamefont {Naber}, \citenamefont {Benson},\ and\
  \citenamefont {Wegener}}]{shiWiringPrecharacterizedSinglephoton2016}%
  \BibitemOpen
  \bibfield  {author} {\bibinfo {author} {\bibfnamefont {Q.}~\bibnamefont
  {Shi}}, \bibinfo {author} {\bibfnamefont {B.}~\bibnamefont {Sontheimer}},
  \bibinfo {author} {\bibfnamefont {N.}~\bibnamefont {Nikolay}}, \bibinfo
  {author} {\bibfnamefont {A.~W.}\ \bibnamefont {Schell}}, \bibinfo {author}
  {\bibfnamefont {J.}~\bibnamefont {Fischer}}, \bibinfo {author} {\bibfnamefont
  {A.}~\bibnamefont {Naber}}, \bibinfo {author} {\bibfnamefont
  {O.}~\bibnamefont {Benson}},\ and\ \bibinfo {author} {\bibfnamefont
  {M.}~\bibnamefont {Wegener}},\ }\bibfield  {title} {\bibinfo {title} {Wiring
  up pre-characterized single-photon emitters by laser lithography},\ }\href
  {https://doi.org/10.1038/srep31135} {\bibfield  {journal} {\bibinfo
  {journal} {Sci Rep}\ }\textbf {\bibinfo {volume} {6}},\ \bibinfo {pages}
  {31135} (\bibinfo {year} {2016})}\BibitemShut {NoStop}%
\end{thebibliography}%
\end{document}